\documentclass[aps,pra,reprint,superscriptaddress]{revtex4-2}
\usepackage{mathtools,amssymb,amsmath,commath}
\usepackage{graphicx}% Include figure files
\usepackage{dcolumn}% Align table columns on decimal point
\usepackage{bm,mathptmx}% bold math
\usepackage[colorlinks=true,bookmarks=true,allcolors=blue,citecolor=red,urlcolor=blue]{hyperref}

\usepackage{tikz}
\DeclareRobustCommand*\circled[1]{\tikz[baseline=(char.base)]{
		\node[shape=circle,draw,inner sep=0.2pt] (char) {#1};}}

\begin{document}
	
	\title{Optical beam propagation inside a graded-index fiber with saturable nonlinearity }
	\author{Tiyas Das}
	\affiliation{Department of Physics, Indian Institute of Technology Kharagpur, Kharagpur, India, 721302.
	}
	\author{Anuj Pratim Lara}
	\affiliation{Department of Physics, Indian Institute of Technology Kharagpur, Kharagpur, India, 721302.
	}
	\author{Samudra Roy }
	\email{samudra.roy@phy.iitkgp.ac.in}
	\affiliation{Department of Physics, Indian Institute of Technology Kharagpur, Kharagpur, India, 721302.
	}	
	\author{Govind P. Agrawal }
\affiliation{The Institute of Optics, University of Rochester, Rochester, NY 14627, USA
	}
	
	\begin{abstract}
We study theoretically the spatial evolution of optical beams inside a graded-index fiber exhibiting saturable nonlinearity. Utilizing an approach based on the variational principle, we identify the existence of bistable spatial solitons inside such a nonlinear medium, whose stability, analyzed through a linear stability analysis, is due to the saturating nature of the nonlinearity. Spatial solitons adhere to a specific amplitude-width relationship. Any deviation from this relationship leads to oscillating-type solutions with a period that increases with saturation level of the nonlinearity. Theoretically calculated values of this period agree well with numerical findings.
	\end{abstract}
	
	\maketitle

\section{Introduction}
	
The propagation of high-intensity beams inside waveguides such as an optical fiber has been studied extensively over the past few decades, both experimentally and theoretically \cite{agrawal2000nonlinear, akhmediev2005dissipative}. One of the fascinating aspects of this work is the formation of spatial solitons in the case of continuous-wave (CW) beams, and temporal solitons in the case of pulsed beams. Such solitons form owing to a subtle balance between the diffraction/dispersion and the Kerr nonlinearity providing an intensity dependent increase in the refractive index of the medium \cite{agrawal2000nonlinear,kivshar2003optical}. In the spatial case, the formation of solitons in planar waveguides was predicted in several studies \cite{Maneuf_1988,Silberberg_1990, Desaix_1991, Snyder_1991,Sammut_1993,Karlsson_1991} and also observed experimentally \cite{Aitchison_1990}. It is known that spatial solitons forming because of the Kerr nonlinearity become unstable in more than one transverse dimensions \cite{Silberberg_1990, Desaix_1991}. To prevent this, the use of saturable nonlinearity has been considered using materials such as photorefractive crystals \cite{Bian_1997, Christodoulides_1995, Lin_2013}, semiconductor-doped glasses \cite{Coutaz_1991}, and negative-index media \cite{Maluckov_2008}.

One way to realize self-guided propagation of optical beams in two transverse dimensions is to make use of graded-index (GRIN) fibers \cite{agr23}. GRIN fibers were fabricated and studied in the 1970s in the context of optical communications \cite{Uchida_1970}. More recently, GRIN fibers have been used for supercontinuum generation \cite{lopez2016visible, Eslami_2022,Krupa_2016}, spatiotemporal mode locking of fiber lasers \cite{Wright_2016, Wright_2017,Qin_2018}, ultrabroadband dispersive radiation \cite{wright2015,Wright_2015}, spatial beam cleanup \cite{guenard_2017, guenard2017kerr}, and high-power fiber amplifiers \cite{Lara_2024,Maity_2024}. One of the advantages of GRIN fibers is that their parabolic refractive-index profile leads to periodic self-imaging of optical beams, launched with an input power less than the critical power associated with the beam collapse induced by self-focusing \cite{Karlsson_1992}.
	
The formation of multimode solitons in GRIN fibers has attracted considerable recent attention \cite{ren13,ahs18,sun24}. However, most earlier work has focused on temporal solitons forming in the presence of the Kerr nonlinearity, as this kind of nonlinearity dominates in silica-based fibers. In this study, we focus on the formation of spatial solitons and consider the propagation of CW beams inside a GRIN fiber doped with nanoparticles exhibiting saturable nonlinearity. Through an analysis based on the variational technique, we discover bistable spatial solitons with a unique amplitude-width relationship, which remain stable even at high powers owing to the  saturable nonlinearity of such GRIN fibers. The stability of bistable solitons under amplitude noise is analyzed through a linear stability analysis. We also examine the formation of similaritons, exhibiting periodic spatial oscillations and occurring when the saturation of nonlinearity is weaker compared to the self-imaging induced by the index gradient.
		
\section{Nonlinear propagation equation}

\begin{figure}[tb!]
  	\includegraphics[width=\linewidth]{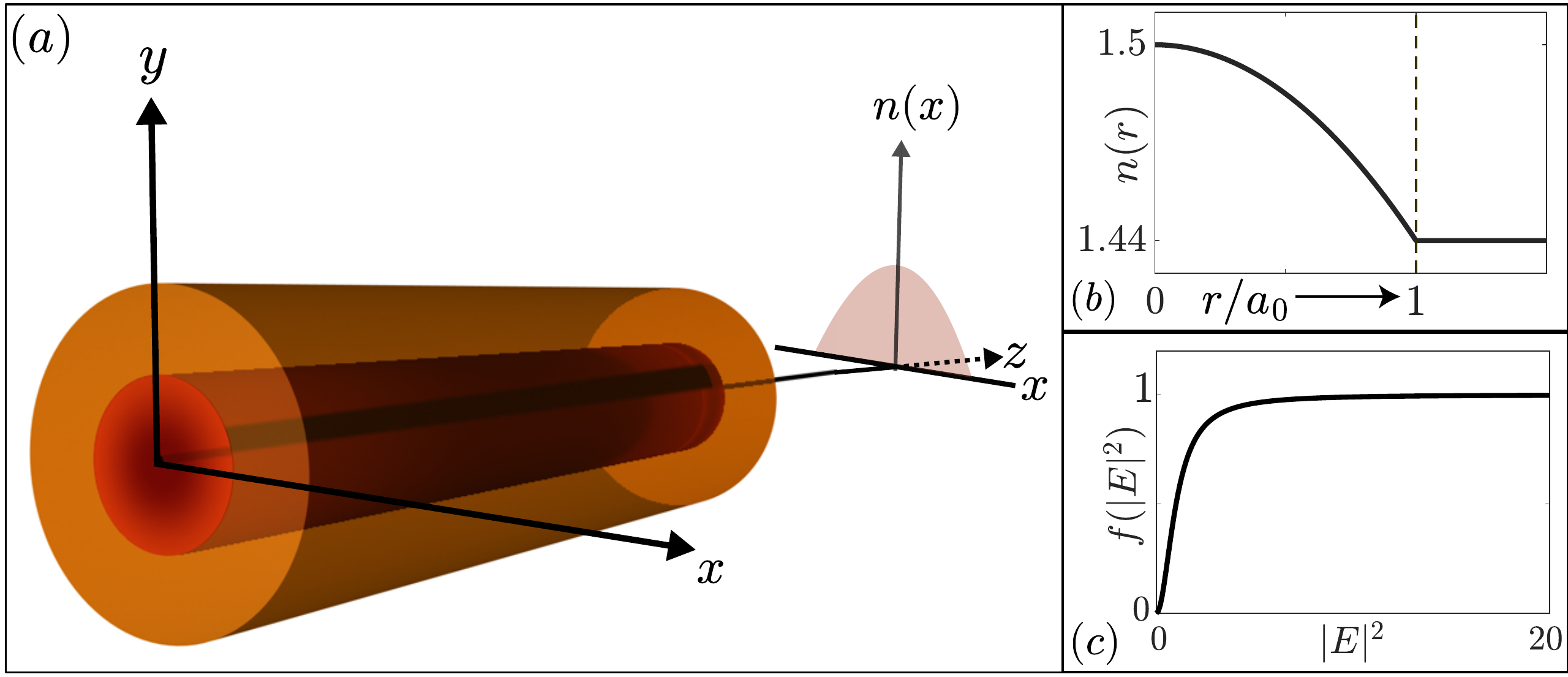}
\caption{($a$) Schematic of a GRIN fiber with ($b$) a parabolic index profile and
  		($c$) a saturable nonlinearity.}
  	\label{Fig1}
  \end{figure}

As illustrated in Fig.~\ref{Fig1}, we consider a GRIN fiber whose core has a parabolic index profile and has been doped with semiconducting nanoparticles to realize a saturable nonlinearity. The refractive index experienced by an incident optical beam with the local intensity $I$ then has the form
	\begin{equation}
		n=n_0\left(1-\frac{1}{2}b^2 r^2\right)+n_2 F(I),
		\label{Eq1}
	\end{equation}
where $r=\sqrt{x^2+y^2}$  is the radial distance from the center of the core and $n_0$ is the refractive index at $r=0$. The index gradient $b$ is defined as $b=a_0^{-1} \sqrt{2\Delta}$, where $a_0$ is the core's radius and $\Delta= (n_0-n_c)/n_0$, $n_c$ being the cladding index. The Kerr coefficient is denoted by $n_2$. The saturable nonlinear response, governed by $F(I)$, has the standard form
    \begin{equation}
    	F(I)=\frac{I}{(1+I/I_{sat})},
    	\label{Eq2}
    \end{equation}
where $I_{sat}$ is the saturation intensity.

A quasi-CW beam, with a narrow spectrum centered at frequency $\omega_0$, is incident on the GRIN fiber shown in Fig.~\ref{Fig1}. Assuming that the polarization of the beam does not change inside the fiber, the evolution of its electric field is governed by the Helmholtz equation:
    \begin{equation}
    	\nabla^{2} E+n^2k^2_{0}E=0,
    	\label{Eq3}
    \end{equation}
where $k_{0}=\omega_{0}/c$ is and $c$ is the velocity of light in free space. Writing the electric field as $E(\vec{r})=A(r,z)e^{ikz}$ and exploiting the paraxial approximation, we obtain
    \begin{equation}
    i\frac{\partial A}{\partial z}+\frac{1}{2k}\nabla^2_{T}A-\frac{1}{2}kb^2 r^2A+ \frac{n_2k_0|A|^2}{(1+|A|^2/|I_{sat})}A=0.
    	\label{Eq4}
    \end{equation}
Here $A(r,z)$ is the slowly varying amplitude of the CW beam and $k=n_0k_0$ is the propagation constant.

For the following analysis, it is useful to normalize Eq.~(\ref{Eq4}) using the dimensionless variables defined as
    \begin{equation}
    \rho= r/w_0, \quad \xi = bz, \quad u = A/\sqrt{I_0},
    \end{equation}
where $w_0 = (kb)^{-1/2}$ and $I_0 = b/(n_2k_0)$. Physically, $w_0$ is the spot size of the fundamental mode associated with a GRIN fiber and $I_0$ sets an intensity scale such that the nonlinear index change satisfies $n_2I_0=b/k_0$. If we focus only on the radially symmetric solutions, Eq.~(\ref{Eq4}) is transformed into the following equation:
    \begin{equation}
    i\frac{\partial u}{\partial\xi} +\frac{1}{2}\left(\frac{\partial^2}{\partial \rho^2} +\frac{1}{\rho}\frac{\partial}{\partial \rho}\right)u-\frac{\rho^2}{2} u +\frac{\mu|u|^2u}{1+s|u|^2}=0,
    	\label{Eq5}
    \end{equation}
where we introduced two dimensionless parameters as
    \begin{equation}
        s=I_0/I_{sat}, \qquad \mu={\rm sgn}(n_2).
    \end{equation}
No saturation of nonlinearity occurs for $s=0$. The two values $\mu=\pm1$ indicate the focusing or defocusing nature of the nonlinearity. We focus on the self-focusing case as it  is more realistic for silica-based GRIN fibers.

We present our theoretical results using normalized parameters. To relate them to values used in an actual experiment, we provide estimates of $w_0$ and $I_0$ for a realistic GRIN fiber designed with $a_0=25~\mu$m, $\Delta=0.005$, and $n_0=1.45$. At wavelengths near 1550~nm where fiber's loss is minimum, we estimate that $b=4$ mm$^{-1}$, $w_0=5~\mu$m, and $I_0=3$ TW/cm$^2$.

\section{Bistable Spatial Solitons}

A spatial soliton is a self-guided beam that maintains its shape and width in space owing to a balance between the diffractive and nonlinear effects. To find them, we seek for a stationary solution in the form $u(\rho, \xi)=f(\rho)e^{iq\xi}$. Substituting this form in Eq.~(\ref{Eq5}), we obtain the following equation for $f(\rho)$:
  \begin{equation}
  	-qf +\frac{1}{2}\left(\frac{d^2}{d\rho^2}+\frac{1}{\rho}\frac{d}{d\rho}\right)f -\frac{\rho^2}{2} f+\frac{\mu f^3}{1+sf^2}=0.
  	\label{Eq6}
  \end{equation}
This nonlinear equation can be solved numerically to find $f(\rho)$. To gain physical insight, we solve it approximately with the \textit{Ribts's optimization} technique and assume that the spatial soliton has a Gaussian shape in the form $f(\rho)= \mathcal{A}\exp[-(\rho^2/2 \mathcal{R}^2)]$, where, $\mathcal{A}$ and $\mathcal{R}$ represent the amplitude and the width of the soliton. They satisfy a specific relation required by Eq.~(\ref{Eq5}).

The Lagrangian density $\mathcal{L}_{D}$ corresponding Eq.~\eqref{Eq6} is found to be
   \begin{equation}
   	\mathcal{L}_{D}= \frac{\rho}{2}\left(q+\frac{\rho}{2}\right)f^2
   	+\frac{\rho}{4}\left(\frac{df}{d\rho}\right)^2-\frac{\mu\rho}{2s^2} [sf^2-\ln(1+sf^2)].
   	\label{Eq7}
   \end{equation}
The Lagrangian, $\mathcal{L}=\int_{0}^{\infty}\mathcal{L}_{D} d\rho$, is obtained by employing the Gaussian ansatz for  $f(\rho)$ and integrating over $\rho$. The result is:
   \begin{equation}   	
   \mathcal{L}=\frac{\mathcal{A}^2}{8}\big(1+2q\mathcal{R}^2+\mathcal{R}^4\big)
   -\frac{\mu\mathcal{R}^2}{4s^2}\big(s\mathcal{A}^2+{\rm Li}_2(-s\mathcal{A}^2)\big),
   	\label{Eq8}
   \end{equation}
where ${\rm Li}_2(x)$ is the dilogarithm function defined as
   \begin{equation}   	
   {\rm Li}_2(x)=-\int_{0}^{x}\frac{\ln(1-t)}{t}dt.
   \end{equation}
Employing the Euler--Lagrange equation, $\partial\mathcal{L}/\partial Q=0$ with $Q = \mathcal{A}$ and $\mathcal{R}$, we obtain the following unique relationship between $\mathcal{A}$ and $\mathcal{R}$:
   \begin{equation}
   	\mathcal{R} =\frac{1}{2\Gamma}\Big(\beta+\sqrt{\beta^2+4\Gamma^2}\Big)
   ^{1/2},
   	\label{Eq9}
   \end{equation}
where $\beta=2[\ln(1+s\Gamma)+{\rm Li}_2(-s\Gamma)]/s^2$ and $\Gamma=\mathcal{A}^2$.

\begin{figure}[tb!]
   	\includegraphics[width=\linewidth]{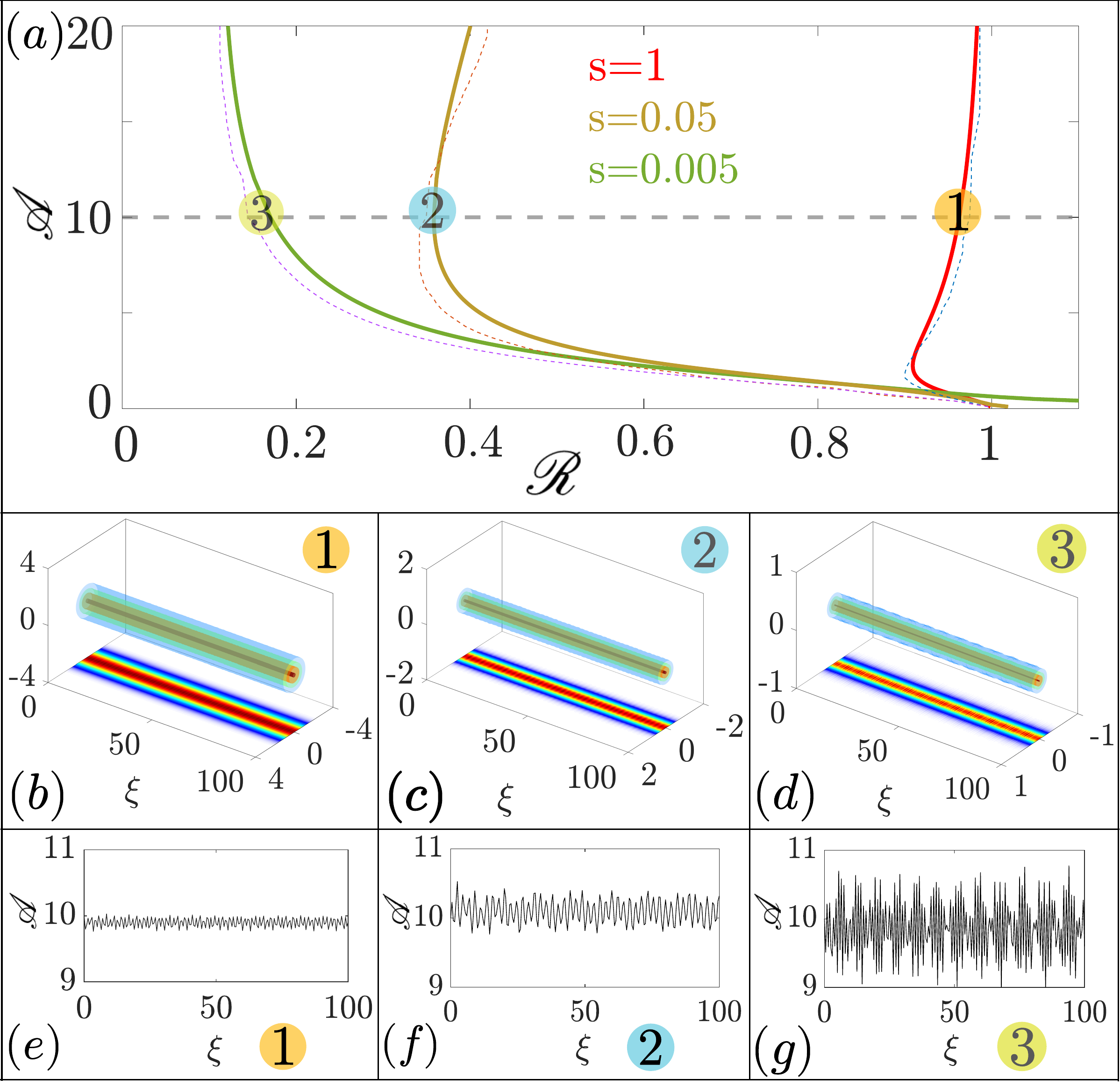}
\caption{($a$) Comparison of analytical (solid lines) and numerical (dashed lines)  curves showing the $\mathcal{A}$-$\mathcal{R}$ relationship for three values of $s$. ($b$)-($d$) Soliton-like propagation of in three cases indicated by circles. ($e$)-($g$) Amplitude variations in the same three cases.}
   	\label{Fig2}
   \end{figure}

The relationship between the soliton's width and amplitude given in Eq.~\eqref{Eq9} is plotted in Fig.~\ref{Fig2}($a$) for three values of the saturation parameter, $s=1,0.05,0.005$ (solid lines). These analytical predictions are verified numerically by solving Eq.~\eqref{Eq6} with the boundary condition $f(\rho)=0$ as $\rho \rightarrow \infty$ (dotted lines). The agreement is quite good for large values of $s$ and remains reasonable even for low values of $s$. This consistency between the analytical and numerical results validates our analytical approach. The most noteworthy feature of Fig.~\ref{Fig2}($a$) is the bistable behavior, allowing for two stable beams of the same width but different amplitudes. It occurs for values of $s>0.01$ and is a known feature of saturable nonlinearity.

The soliton-like propagation of a Gaussian beam is shown in parts ($b$)--($d$) of Fig.~\ref{Fig2} by solving Eq.~\eqref{Eq6} numerically in three cases marked by circles in part ($a$). The soliton's amplitude does not remain constant and exhibits small variations depicted in parts ($e$)-($g$). As seen there, amplitude variations are relatively negligible for $s=1$, but they increase rapidly as $s$ decreases. Recalling that $s$ is a measure of the degree of nonlinearity saturation, it is not surprising that spatial solitons become unstable when $s<0.01$ because the saturation of nonlinearity becomes negligible. It is known that a GRIN fiber with unsaturable Kerr nonlinearity does not support the formation of stable spatial solitons.

\section{Stability analysis}

The stability of spatial solitons governed by Eq.~\eqref{Eq5} can be examined mathematically using a technique known as the linear stability analysis (LSA). Any stationary solution of this equation, represented by $f(\rho)$, is perturbed by a small amount that can be generally complex. The perturbed optical field is written as
    \begin{equation}
    u(\xi,\rho) = \left[f(\rho)+v(\rho)e^{\Lambda\xi}
        + w^*(\rho)e^{\Lambda^*\xi}\right]e^{iq\xi},
    \end{equation}
where $v(\rho)$ and $w(\rho)$ both can grow exponentially with a growth rate set by $\Lambda$. By substituting this form of $u(\xi,\rho)$ into Eq.~\eqref{Eq5} and neglecting all terms higher than the first order in $v$ and $w$ , we obtain two linear equations for $v$ and $w$ that can be solved to find the growth rate $\Lambda$. The two linear equations are combined into a matrix form and written as the eigenvalue problem $\mathcal{M}X=\Lambda X$, where $X=[v,w]^T$ is a column vector and the matrix $\mathcal{M}$ has the form
    \begin{equation}
    	\mathcal{M} = i\begin{bmatrix}
    		\tilde{\nabla}+\alpha_0 & \alpha_1 \\
    		-\alpha_1 & -(\tilde{\nabla}+\alpha_0)
    	\end{bmatrix}.
    \end{equation}
Here $\tilde{\nabla}=\frac{1}{2}(\partial^2_x+\partial^2_y)$ is a differential operator and
\begin{equation}
    \alpha_0=-q-\frac{1}{2}(x^2+y^2)+\mu f^2\mathcal{G}(1+\mathcal{G}), \quad \alpha_1=\mu f^2\mathcal{G}^2,
\end{equation}
with $\mathcal{G}=(1+sf^2)^{-1}$.

The Fourier collocation method \cite{yang2010nonlinear} was employed to analyze the eigenvalue spectrum of the operator $\mathcal{M}$. An eigenvalue $\Lambda$ with a positive real part ($\Re(\Lambda)>0$) indicates instability because of the exponential growth of any perturbation, while any eigenvalue with $\Re(\Lambda)\leq 0$ indicates stability because all perturbations decay exponentially with propagation. The case $\Re(\Lambda)=0$ for any eigenvalue corresponds to neutral stability. As that eigenvalue is purely imaginary, perturbations evolve in a periodic fashion rather than decaying.

\begin{figure}[tb!]
    	\includegraphics[width=\linewidth]{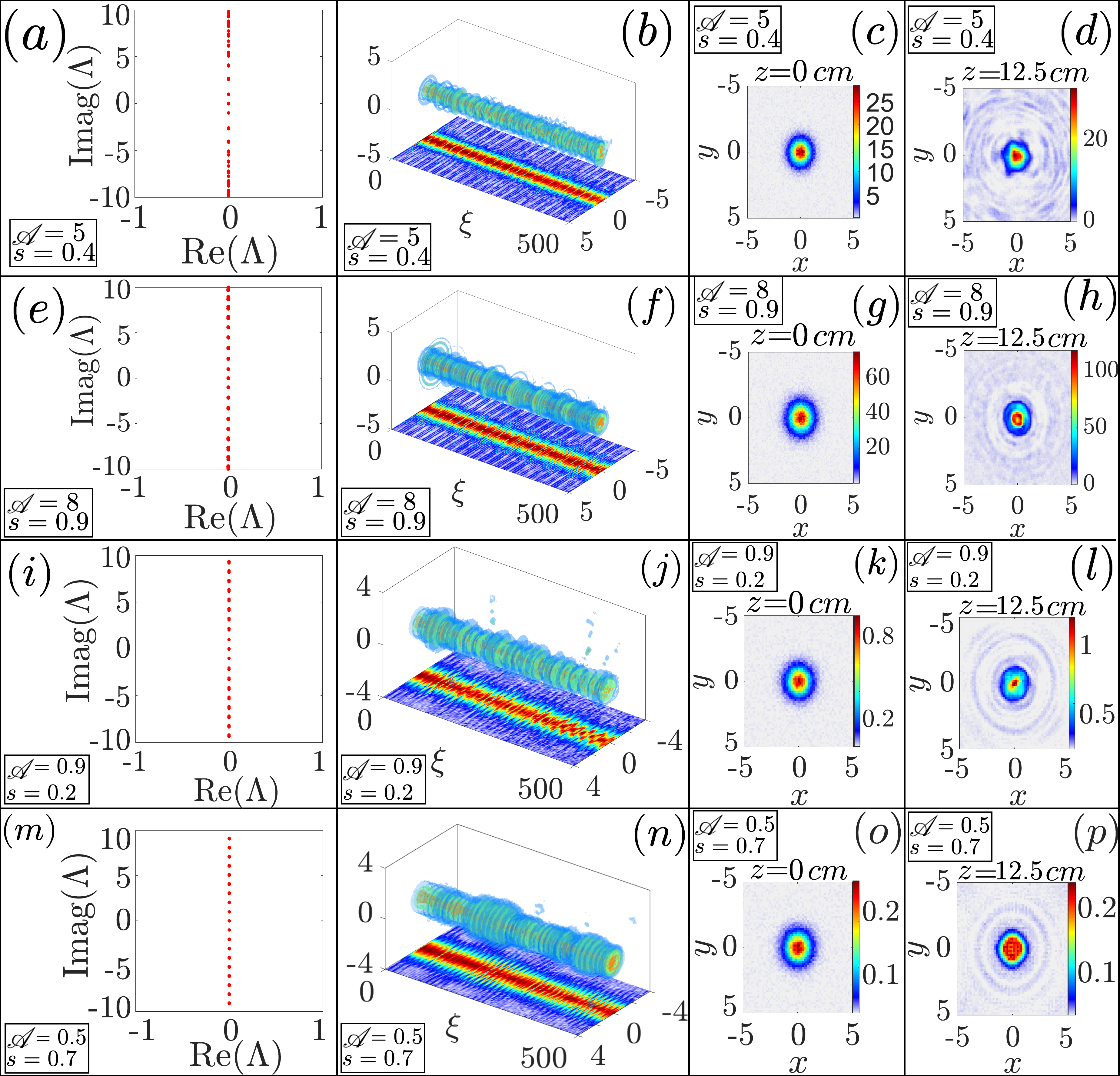}
\caption{Stability features of bistable solitons for four different combinations of $\mathcal{A}$ and $s$. Top two rows correspond to the upper bistable branch, and the bottom two rows to the lower bistable branch. In each case, the first column shows the real and imaginary parts of the eigenvalues and the second column shows the evolution of a perturbed soliton. The third and fourth columns show spatial profiles of the soliton at the input and output ends of the GRIN fiber.} \label{Fig5}
    \end{figure}

To reveal how the stability depends on the parameters $s$ and $\mathcal{A}$, Figure \ref{Fig5} show various stability features of bistable solitons for four different combinations of the parameters $\mathcal{A}$ and $s$. Top two rows correspond to the upper bistable branch, and the bottom two rows to the lower bistable branch. In each case, the first column shows the real and imaginary parts of the eigenvalues calculated numerically. It reveals that the real part of $\lambda$ is zero for all eigenvalues, indicating neutral stability with an oscillatory decay of perturbations imposed on the Gaussian beam propagating as a spatial soliton.
		
The evolution of the Gaussian beam inside the GRIN fiber in the same four cases is shown in the second column of Figure \ref{Fig5}. The initial amplitude of the beam was perturbed at the input end of the by adding random noise (maximum amplitude 10\%). The corresponding input and output shapes are shown in the third and fourth columns, respectively. The output beam in all cases show circular rings that correspond to a periodic evolution of the perturbation because of purely imaginary eigenvalues seen in the first column.
		
\section{Spatial Similaritons}

Spatial solitons found in Section~3 maintain their initial width inside the GRIN fiber. In this section, we extends this analysis by allowing variations in both the amplitude ($\mathcal{A}$) and width ($\mathcal{R}$) of the beam along the fiber's length. Such solutions are sometimes referred to as spatial similaritons \cite{pon06, pon07}.

The starting point is Eq.~(\ref{Eq5}), which we solve approximately with the variational technique. We still assume a Gaussian shape of the beam but allow for changes in all parameters of the beam, including curvature of the phase front, through the variational ansatz
  \begin{equation}
  	 u(\rho,\xi)=\mathcal{A}(\xi)\exp[-\rho^2/2 \mathcal{R}^2(\xi) +i\phi(\xi)
        +id(\xi)\rho^2].
  	 \label{Eq11}
  \end{equation}

The Lagrangian density $\mathcal{L}_{D}$ corresponding to Eq.~(\ref{Eq5}) is
  \begin{equation}
  	\begin{split}
  	\mathcal{L}_{D} = \frac{i\rho}{2}\left(u \frac{d u^*}{d\xi}-u^{*}\frac{du} {d\xi}\right)+\frac{\rho}{2}\left|\frac{\partial u}{\partial\rho}\right|^2 +\frac{\rho^3}{2}|u|^2\\
  -\frac{\mu\rho}{s^2}[s|u|^2- \ln(1+s|u|^2)].
  	\end{split}
  	\label{Eq10}
  \end{equation}
The Lagrangian is obtained by using the Gaussian ansatz and carrying out the integration $\mathcal{L}=\int_{0}^{\infty} 	\mathcal{L}_{D} d\rho$. The result is found to be
 \begin{equation}
 	\begin{split}
 		\mathcal{L}= \frac{\mathcal{A}^2\mathcal{R}^2}{2}\left(\phi_{\xi}
    + d_{\xi}\mathcal{R}^2 \right)+\frac{\mathcal{A}^2}{4}\left(1+ 4d^2
        \mathcal{R}^4 +\mathcal{R}^4 \right)\\
    -\frac{\mu \mathcal{R}^2}{2s^2}[s\mathcal{A}^2+{\rm Li}_2(-s\mathcal{A}^2)],
 		\end{split}
 	\label{Eq12}
 \end{equation}
where $\phi_\xi$ and $d_{\xi}$ denote a derivative with respect to $\xi$.

Using the Euler--Lagrange equation, $\frac{\partial}{\partial\xi}\left( \frac{\partial\mathcal{L}} {\partial Q_{\xi}}\right)-\frac{\partial\mathcal{L}}{\partial Q}=0$, with $Q = \mathcal{A}$, $\mathcal{R}$, $d$ and $\phi$, we obtain the following four coupled differential equations governing the evolution of beam parameters:
 \begin{subequations}
 	\begin{equation}
 		\frac{\partial \mathcal{A}}{\partial \xi}=-2d\mathcal{A}
 		\label{Eq13a}
 	\end{equation}
 	\begin{equation}
 		\frac{\partial \mathcal{R}}{\partial \xi}=2d\mathcal{R}
 		\label{Eq13b}
 	\end{equation}
 	\begin{equation}
 		\begin{split}
 		\frac{\partial d}{\partial \xi}=-2d^2-\frac{1}{2}
    +\frac{1}{2\mathcal{R}^4} +\frac{\mu}{s^2\mathcal{A}^2\mathcal{R}^2} \\
    \times [\ln(1+s\mathcal{A}^2)+{\rm Li}_2(-s\mathcal{A}^2)]
 		\end{split}
 		\label{Eq13c}
 	\end{equation}
 	\begin{equation}
 		\begin{split}
 		\frac{\partial \phi}{\partial \xi}=\frac{\mu}{s}
    -\frac{1}{\mathcal{R}^2} -\frac{\mu}{s^2\mathcal{A}^2} \\
    \times [2\ln(1+s\mathcal{A}^2)+{\rm Li}_2(-s\mathcal{A}^2)].
 		\end{split}
 		\label{Eq13d}
 	\end{equation}
 \end{subequations}

\begin{figure}[tb!]
 	\includegraphics[width=\linewidth]{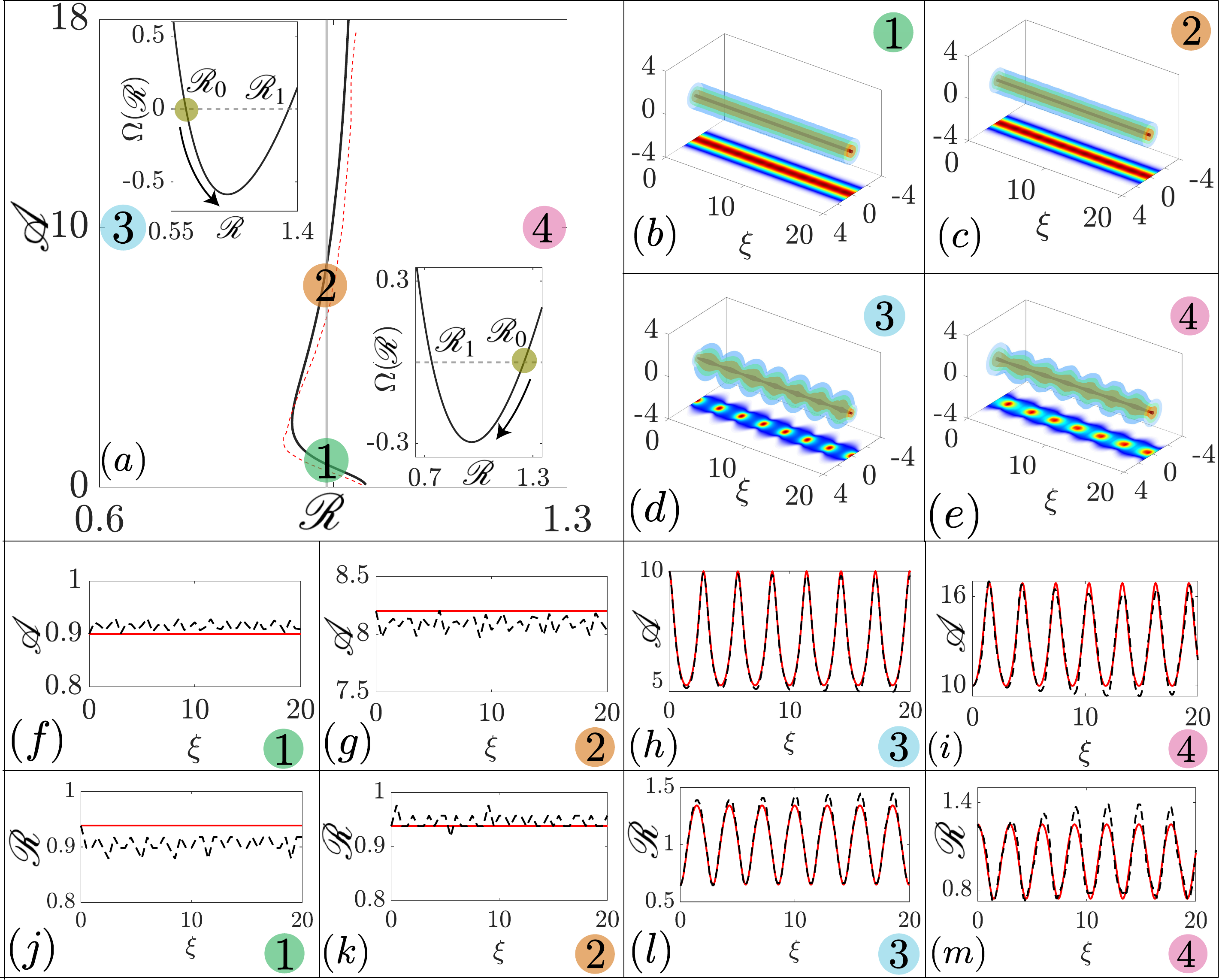}
\caption{Bistable $\mathcal{A}$-$\mathcal{R}$ curve obtained analytically (solid line) and numerically (dotted line) using $s=0.8$. Points \circled{1} and \circled{2} mark two amplitudes allowed for a specific width of the soliton. Two insets shows the potential that governs beam-width oscillations of similaritons at points \circled{3} and \circled{4}. ($b$)-($e$) Evolution of Gaussian-beams at these four points. ($f$)-($k$) Periodic variations of beam's amplitudes and widths in the four cases. Solid lines correspond to variational results and dotted lines show numerical data.} \label{Fig3}
 \end{figure}

We solve these equations numerically for a specific case of $s=0.8$, and the the results are shown in Fig.~\ref{Fig3}. The results of Section~3 corresponding to spatial solitons are recovered by setting all $\xi$ derivatives to zero. In this case, Part $a$ shows the relationship between the amplitude ($\mathcal{A}$) and width ($\mathcal{R}$) of the spatial soliton using both the variational analysis (solid line) and numerical simulations (dotted line). We choose four points, labeled \circled{1} through \circled{4}, such that the first two lie on the solid curve and produce spatial solitons, while the last two are far away from this curve and form similaritons whose widths and amplitudes vary with propagation.

To reveal the differences in these two situations, we solve Eq.~(\ref{Eq5}) numerically with a Gaussian beam launched with parameters corresponding the four  points, labeled \circled{1}--\circled{4}, and show the beam's evolution in parts ($b$)-($e$). Further, parts ($f$)-($m$) also depict the amplitude and width variations obtained using the variational results in Eqs.\eqref{Eq13a}--\eqref{Eq13d}), showing good agreement with numerical data. For points \circled{1} and \circled{2}, both the amplitude and width remain constant during propagation, consistent with theoretical predictions. Conversely, spatial similaritons at points \circled{3} and \circled{4}) exhibit a breathing behavior such that the amplitude and width vary in a periodic fashion along the fiber's length. We have ensured that the total energy,
  \begin{equation}
        \mathcal{E}_0=\int_{0}^{\infty} |u(\rho,\xi)|^2 \rho d\rho=\mathcal{A}^2(\xi)\mathcal{R}^2(\xi)/2,
  \end{equation}
of the beam remains conserved all along the fiber.

We can estimate the period of width oscillations by exploiting Eq.~\eqref{Eq13b} and Eq.~\eqref{Eq13c}. Taking a second derivative of Eq.~\eqref{Eq13b}, it is possible to construct the following second-order differential equation for $\mathcal{R}$:
  \begin{equation}
  	\frac{d^2\mathcal{R}}{d\xi^2}+\mathcal{R}=\frac{1}{\mathcal{R}^3}+\frac{\mu \mathcal{R}}{\mathcal{E}_0s^2}\left[\ln\left(1+\frac{2s\mathcal{E}_0}{\mathcal{R}^2}
  \right)+{\rm Li}_2\left(\frac{-2s\mathcal{E}_0}{\mathcal{R}^2}\right)\right].
  	\label{Eq15}
  \end{equation}
By integrating this equation once, we can write the result as
   \begin{equation}
   	\left(\frac{d\mathcal{R}}{d\xi}\right)^2+\Omega(\mathcal{R})=0,
   	\label{Eq16}
   \end{equation}
where $\Omega(\mathcal{R})$ acts as a potential function and has the form
   \begin{equation}
   	\begin{split}   	
   \Omega(\mathcal{R})= (\mathcal{R}^2-\mathcal{R}^2_0)
   +\left(\frac{1}{\mathcal{R}^2}-\frac{1}{\mathcal{R}^2_0}\right)\\
   +\frac{\mu}{\mathcal{E}_0s^2}\left[\mathcal{R}^2_0{\rm Li}_2 \left(\frac{-2s\mathcal{E}_0}{\mathcal{R}^2_0}\right)
   -\mathcal{R}^2{\rm Li}_2\left(\frac{-2s\mathcal{E}_0}{\mathcal{R}^2}\right)\right],
   	\end{split}
   	\label{Eq17}
   \end{equation}
and $\mathcal{R}_0$ denotes the initial width of the beam.

Equation \eqref{Eq16} can be viewed as describing the motion of a point mass under the potential $\Omega(\mathcal{R})$ \cite{karlsson1992optical, Ianetz_2013}. This potential vanishes at two values of $\mathcal{R}$, denoted by $\mathcal{R}_0$ and $\mathcal{R}_1$. The beam's width oscillates between these two values. If $\mathcal{R}_0 < \mathcal{R}_1$, the width first increases and then returns to its initial state, as illustrated for point \circled{3} in Fig.~\ref{Fig3}($l$). Conversely, when $\mathcal{R}_0 > \mathcal{R}_1$, a focusing-type oscillation occurs, characterized by an initial decrease in beam's width before returning to its original value, as depicted in Fig.~\ref{Fig3}($m$). The two insets in Fig.~\ref{Fig3}(a) show the potential function for both cases with the initial beam-width indicated by a circular spot. A spatial cases corresponds to the condition $d\Omega(\mathcal{R})/d\mathcal{R}=0$ for a beam launched with the width $\mathcal{R}_w$ such that the grin-induced focusing and diffraction balance each other. For a given energy, $\mathcal{R}_w$ is the constant width predicted by Eq.~\eqref{Eq9}.

The potential concept can be further exploited for calculating the breathing period $w_p$ by using $\mathcal{R}=\mathcal{R}_w+\Delta \mathcal{R}$, where $\Delta \mathcal{R}$ measures the change from the stationary width. Inserting this form of $\mathcal{R}$ in Eq.\ (\ref{Eq15}) and linearizing it, we obtain
\begin{equation}
	\frac{d^2(\Delta\mathcal{R})}{d\xi^2}+w_p^2\Delta\mathcal{R}=0,
	\label{Eq28}
\end{equation}
where the breathing period $w_p$ is defined~\cite{Skarka1997-co}
    \begin{equation}
    	w_p =\sqrt{\frac{1}{2}\frac{d^2\Omega(\mathcal{R})}{d\mathcal{R}^2}} \bigg|_{\mathcal{R}=\mathcal{R}_w}.
    	\label{Eq18}
    \end{equation}
Using Eq.~\eqref{Eq17}, $w_p$ is found to be
    \begin{equation} 	
   	w_p = \Big[1+\frac{3}{\mathcal{R}_w^4}-\frac{\mu}{s^2\mathcal{E}_0}
   \Big({\rm Li}_2(-m)+3\ln(1+m)-\frac{2m}{1+m}\Big)\Big]^{1/2},
   	\label{Eq19}
   \end{equation}
where $m=s\mathcal{E}_0/2\mathcal{R}^2_w$.

\begin{figure}[tb!]
  	\includegraphics[width=\linewidth]{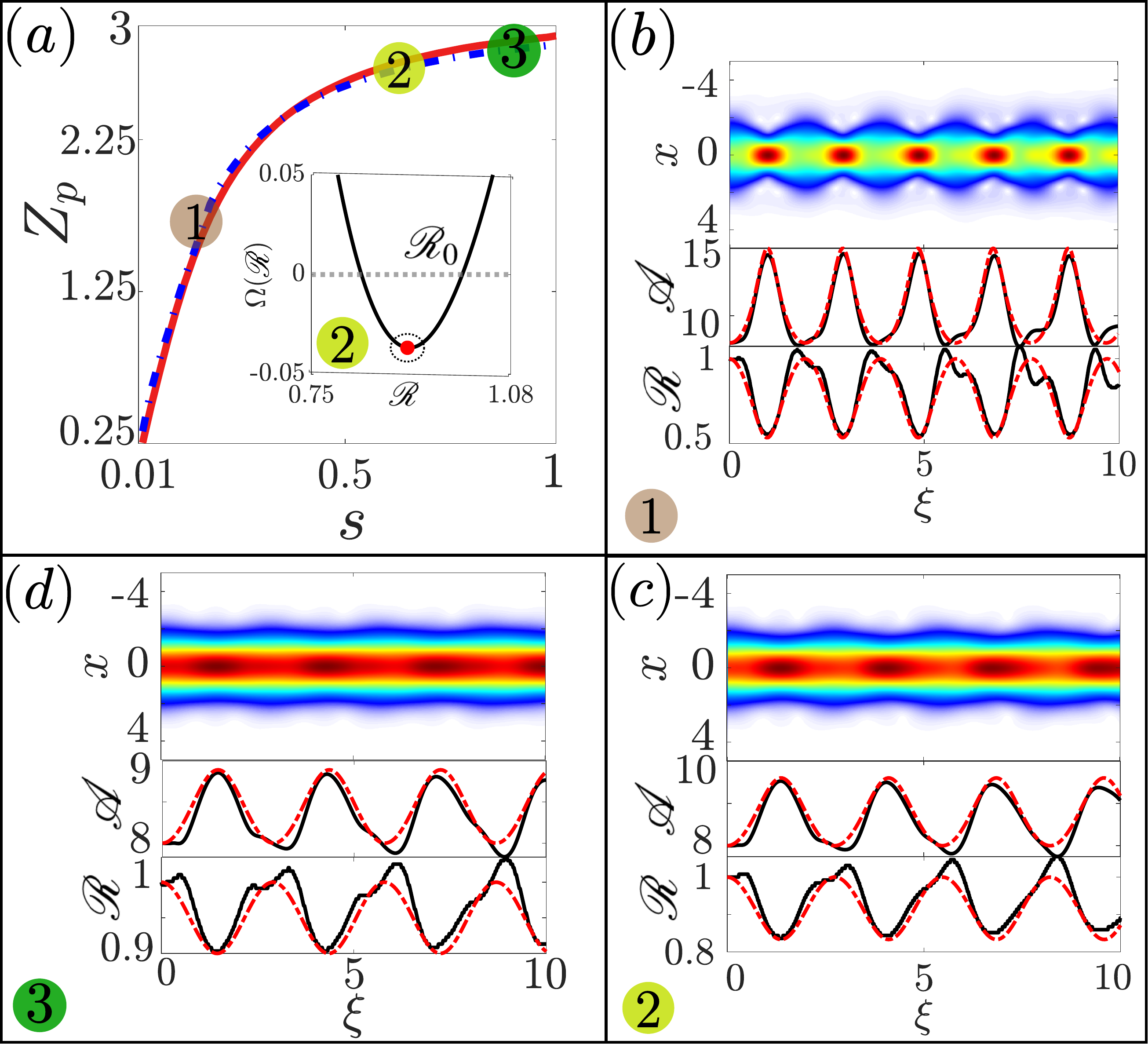}
\caption{Changes in $Z_p$ as a function of $s$ (solid line) predicted by the variational result in Eq.\eqref{Eq18}. The corresponding numerical results are shown by a dotted line for $\mathcal{A}_0=8$ and $\mathcal{R}_0=1$. ($b$)-($d$) Evolution of a Gaussian beam corresponding to points\circled{1}-\circled{3}. Periodic changes in the beam's width ($\mathcal{R}$) and amplitude ($\mathcal{A}$) are also shown as a function of $\xi$. In all cases, the solid and dotted lines correspond to numerical and variational results, respectively.}
  	\label{Fig4}
  \end{figure}

The length $Z_p$ over which one oscillation of the width occurs is given by $Z_p= 2\pi/w_p$. The part ($a$) of Fig.~\ref{Fig4} shows how $Z_p$ varies with the saturation parameter $s$. The solid line corresponds to the analytical results based on Eq.~\eqref{Eq19}, while the dots represent numerical results obtained for the parameter values $\mathcal{A}_0=8$ and $\mathcal{R}_0=1$. Parts ($b$)-($d$) display the beam's evolution together with variations in the amplitude and width of the beam for three different values of $s$, indicated by \circled{1} to \circled{3}. These results agree well with the variational predictions shows by dashed lines. The inset in part ($a$) shows the potential function $\Omega(\mathcal{R})$ at point \circled{2}, indicating that the incident beam undergoes a GRIN-induced focusing-type behavior.

\section{Concluding Remarks}

In this work, we have used the analytical and numerical techniques to study the propagation of optical beams inside a GRIN fiber exhibiting saturation of the Kerr nonlinearity at high intensity levels. We found that spatial solitons, which become unstable in a pure Kerr medium (no saturation of nonlinearity), can form in such a GRIN-type medium. We used a variational technique to derive the relation between the amplitude and width of spatial solitons. We used this relation to demonstrate the existence of bistable spatial solitons of different amplitudes with the same width. Numerical results confirmed the formation of such bistable spatial solitons, but their stability depends on the level of saturation occurring at the peak intensity of the solitons.

To understand fully the nature of stability, we carried out a linear stability analysis and found numerically the eigenvalues that determine whether any perturbation will decay or grow with propagation inside the GRIN fiber. Our results revealed that the stability of spatial solitons depends on the saturation parameter $s$. When $s$ exceeds a value near 0.1, spatial solitons are found to be neutrally stable on both the lower and upper branches of the bistable curve. As the eigenvalues are purely imaginary, any perturbation evolves in a periodic fashion, rather than decaying exponentially along the fiber's length.

We also analyzed what happens when a Gaussian beam is launched such that it does not satisfy the specific amplitude-width relationship required by spatial solitons. We found that spatial similaritons can still form, whose amplitude and width oscillate in a periodic fashion along the fiber's length. This is not surprising in view that GRIN fibers exhibit periodic self-imaging even in the absence of the nonlinear effects. Our analysis shows that the period of width oscillations is influenced by the saturation of nonlinearity such that the period increases rapidly with increasing $s$, before saturating when the value of $s$ approaches 1. The main conclusion of this work is as follows. Unlike a pure Kerr nonlinearity, which does not allow for formation of stable spatial solutions inside GRIN fibers, saturation of the Kerr nonlinearity at high intensity levels permits the formation of stable spatial solitons when a Gaussian beam is launched with a specific amplitude for a given width. When the amplitude and width are chosen arbitrarily, periodically evolving spatial similaritons can still form.
\bibliography{myref}
\end{document}